  \documentclass[preprint]{aastex631}   

\received{October 31, 2023}
\revised{January 22, 2024}
\accepted{February 6, 2024}

\submitjournal{\em Planetary Science Journal}

\usepackage{geometry}      
\usepackage{graphicx}     
 \usepackage{longtable}   
 \usepackage{natbib}       
\usepackage{xcolor}         
\usepackage[normalem]{ulem}
\usepackage{amssymb}   
\usepackage{amsmath}   

\usepackage{epstopdf}            
\usepackage{boxedminipage}  
\usepackage{booktabs}          
\usepackage{topcapt}           

\usepackage{wasysym}     

\graphicspath{{./}{figures/}}   

\begin{document}

\title{Noble Gas Planetology and the Xenon Clouds of Uranus}

\author[0000-0002-2462-4358]{Kevin Zahnle}
 \affiliation{NASA Ames Research Center\\
            Mail Stop 245-3 \\
             Moffett Field, CA 94043, USA}

\begin{abstract}  

Noble gases provide tracers of cosmic provenance that are accessible to a future Uranus Atmospheric Probe.
Argon and krypton are expected to be well-mixed on Uranus with respect to H$_2$ and He, although
condensation at the winter pole may be possible.
The Ar/H$_2$ and Ar/Kr ratios address whether the materials accreted by
Uranus resembled the extremely cold materials accreted by Jupiter's atmosphere, or whether
they were warmer like comet 67P/Churyumov-Gerasimenko, or whether Uranus is like neither.
Xenon condenses as an ice, probably on methane ice, in Uranus's upper troposphere. 
Condensation may complicate interpretation of Xe/H$_2$, 
but it also presents an opportunity to collect concentrated xenon samples suitable for measuring isotopes.
Solar System Xe tracks three distinct nucleosynthetic xenon reservoirs,
one evident in the Sun and in chondritic meteorites, a second in refractory presolar grains, and a third evident in comet 67P/C-G and in Earth's air.
The first and third reservoirs appear to have been captured from different clouds of gas. 
The two gases do not appear to have been well-mixed; moreover, the
 high $^{129}$Xe/$^{132}$Xe ratio in 67P/C-G implies that the gas was captured before the
 initial nucleosynthetic complement of $^{129}$I (15.7 Myr half-life) had decayed.
 Xenon's isotopic peculiarities, if seen in Uranus, could usefully upset our understanding of planetary origins.   
 Krypton's isotopic anomalies are more subtle and may prove hard to measure. 
There is a slight chance that neon and helium fractionations can be used to constrain how Uranus acquired its nebular envelope.

\end{abstract}


\clearpage 

\section{Introduction}
\label{sec:Introduction}

 Noble gases play an outsized role in comparative planetology.
 Their value stems from their volatility and chemical inertness:
 the vast majority of a planet's noble gases are likely to be found in its atmosphere,
and hence are relatively accessible to spacecraft measurements, while their 
 simple geochemistry gives hope that we can understand what they are doing.
Moreover, noble gases often display big signals, both in relative elemental abundances and in isotopic structures
within an element.  

Any Uranus Atmospheric Probe worth flying will attempt to obtain elemental and isotopic abundances of the noble gases.
Noble gases provide most of the spacecraft-accessible tracers of the materials Uranus was made from. 
In particular, noble gas elemental abundances can tell us if Uranus's atmosphere accreted from extremely cold ices
 similar to what we see as vapor in
 Jupiter's atmosphere, or whether it accreted from warmer materials akin to the comet 67P/Churyumov-Gerasimenko, 
 or whether it accreted from something else again \citep{Mandt2020}. 
 Noble gases can also tell us about the history, composition, and evolution of the solar nebula \citep{Mandt2020}.
 Xenon's isotopes in particular can tell us if Uranus accreted from the same mix of interstellar gases as the Sun or if, like Earth's air or 67P/C-G, it records a different provenance.
 On a bigger scale, noble gases provide a test of the convenient metallicity hypothesis---the hypothesis that all elements
 heavier than helium are equally enriched in gas giants---by assessing the ``metals'' in the cold gas giants of our own Solar System.
 Is uniform metallicity a general property of gas giants, including exoplanets, or do we extrapolate too much from Jupiter? 
 
 \medskip

This essay opens with a discussion of how Jupiter's atmophiles and the composition of 67P/C-G can be used to 
predict the noble gases of Uranus.
We next address the heretofore overlooked likelihood that xenon condenses on Uranus.
Third, we review the strangely huge isotopic anomalies seen in solar system xenon
and how the different kinds of Xe are traced back to 
genetically distinct pools of cosmic gas. 
In two brief appendices, we ask if neon could be fractionated by nebular escape, and 
we provide a heuristic derivation of U-Xe,  the original base composition of the Xe in Earth's atmosphere.
U-Xe provides a link between Earth's air and the outer solar system. 

\medskip
\section{Uranian noble gases: two exemplary cases}

\subsection{Uranus inspired by Jupiter}
\label{Jupiter}

\begin{figure}[htbp] 
   \centering
  \includegraphics[width=5in]{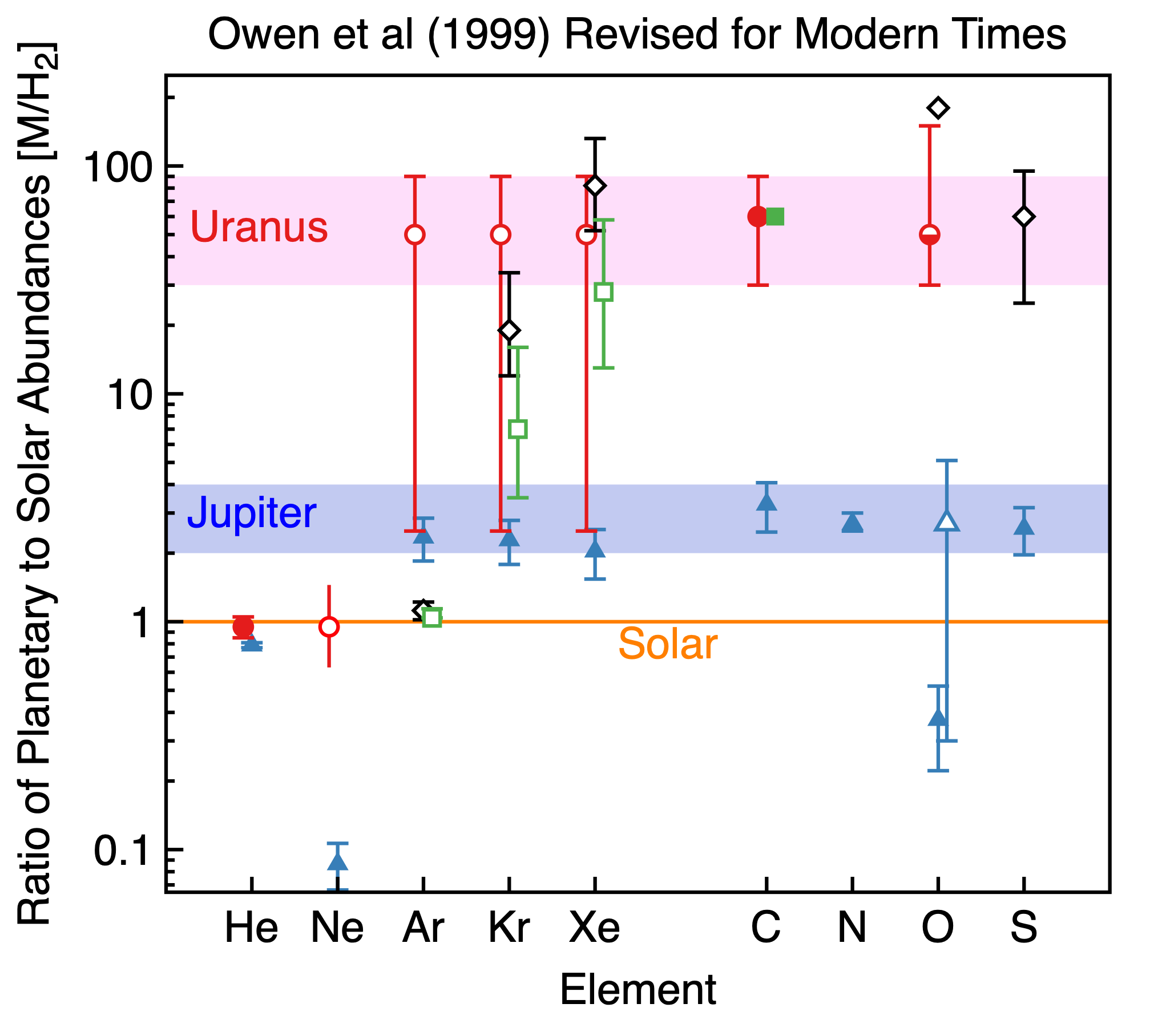} 
   \caption{Imagining noble gas abundances of Uranus.
   Jovian data (filled blue triangles) have been updated from \citet{vonZahn1996}, \citet{Mahaffy2000}, and \citet{Wong2004}.
   The data are plotted as ratios to solar abundances \citep{Lodders2020}. 
   For the jovian N/H$_2$ ratio, we use an NH$_3$ abundance of $340^{+27}_{-17}$ ppmv \citep{Moeckel2023}, 
   to which we add the putative NH$_4$SH cloud using H$_2$S as reported by the Galileo Probe \citep{Wong2004}.
   The open blue triangle is a Juno inference of equatorial H$_2$O \citep{Li2020}.
   The Uranian C/H$_2$ ratio is from CH$_4$ (red filled disk), which presents variable 
   deep atmospheric volume mixing ratios of order 2-5\% \citep{Sromovsky2011}, 
    corresponding to a metallicity enhancement of $60\pm 25$.
   An O/H$_2$ ratio (red half-filled disk) is inferred from thermochemical quenching theory \citep{Venot2020}.
   One set of hypothetical Uranian abundances of Ar, Kr, and Xe is scaled with C from Jupiter (open red circles).
   Another set is scaled from abundances in comet 67P/Churyumov-Gerasimenko assuming a solar C/O ratio (open green squares).
   A third set is scaled from volatile C abundances in 67P/C-G (open black diamonds, using data tabulated by \citet{Mandt2020}).
   In this case we have arbitrarily plotted S but not N.
   For clarity, points are plotted with small artificial shifts along the x-axis.   
}
   \label{fig:one}
\end{figure}

NASA's Galileo Probe found that Jupiter’s atmosphere contains six of the most volatile elements---C, N, S, Ar, Kr, and Xe---at 
abundances that are roughly $2.5\times$ what they are in a 
solar composition gas \citep{Owen1999, Mahaffy2000,Atreya2003,Wong2004,Mandt2020}. 
The Jovian data normalized to solar composition are shown in Figure \ref{fig:one}.
Carbon, nitrogen, and sulfur are chemically active and form a great variety of compounds that can be rather abundant in volatile-rich meteorites.
Argon, krypton, and xenon are noble gases that do no such thing.
The plausible way to get a nearly uniform enhancement of three chemically active volatile elements and three chemically inactive volatile elements
is to quantitatively freeze out all six as ices \citep{Owen1999}. 
\citet{Owen1999} therefore proposed that there must have existed, and may still exist, a significant population of bodies in the solar system  that condensed at temperatures cold enough to freeze out all the elements other than H, He, and Ne.
The cold bodies need not have been large.  
Snowballs or pebbles could suffice,
provided only that the solids were big enough to separate aerodynamically from the gas \citep{Cuzzi2004,Guillot2006}.
Whatever form the cold matter took, the mass involved must have been enormous to make a major contribution to a planet as big as Jupiter.
Hence a reasonable null hypothesis is to expect a similar noble gas pattern on Uranus, 
scaled up from Jupiter in proportion to the observed abundance of methane.  

With these considerations in mind, we have
updated \citet{Owen1999}'s plot of Jovian C, N, S, Ar, Kr, and Xe abundances
and extended it to encompass Uranus.
These are the subjects of Figure \ref{fig:one}.
The best available data for Uranus are for He and CH$_4$,
as both are abundant enough to strongly affect the atmospheric scale height, which is the quantity that
was measured \citep{Lindal1987,Sromovsky2011}.
Because the scale height depends on the ratio of temperature to mean molecular weight, 
the retrieved He and CH$_4$ abundances are degenerate with temperature.
Retrieval therefore exploits methane condensation as a temperature constraint.
\citet{Sromovsky2011}'s preferred deep methane abundance of 2.3-4\% is consistent with 
highly variable (because methane condenses) telescopic measurements of mixing ratios of 1-5\% \citep{Moses2020}.
Assuming a solar He/H ratio, methane concentrations of 2-5\% correspond to C/H$_2$ enhancements of $60\pm 25$.
The corresponding He/H$_2$ volume mixing ratio is $15\pm 1\%$.

Possible Uranian Ar, Kr, and Xe abundances are scaled from Jupiter 
at a nominal enhancement of $50\times$, but over a very wide range extending from the small $2.5\times$ jovian amplification to
an upper bound of $100\times$.
We have not scaled NH$_3$ or H$_2$S from Jupiter, as neither behave simply enough at Uranian temperatures to yield
trustworthy bulk abundances from remote observations \citep{Guillot2019}.
Nitrogen and sulfur will be targets for {\em in situ} measurements with the Uranus Probe.

Oxygen is not among the six super-volatile elements---H$_2$O is much too refractory. 
The Galileo Probe saw relatively little water \citep{Wong2004}.
Juno microwave retrievals of H$_2$O at Jupiter's equator,
where the NH$_3$ abundance is high and relatively uniform, suggest
that H$_2$O may be amplified above solar by a factor $2.7^{+2.4}_{-1.7}$ \citep{Li2020}.
This new datum is plotted as an open triangle on Figure \ref{fig:one}. 
Recent theoretical thermochemistry quenching calculations suggest that Uranian O/H is enhanced some $45\times$ above the solar ratio
\citep{Venot2020}, which revises a previous estimate of $150\times$ obtained by the same group \citep{Cavalie2017}.
We have plotted Uranian O/H$_2$ on Figure \ref{fig:one} at $50\times$ solar with generous uncertainties.

\medskip

\begin{figure}[htb] 
   \centering
   \includegraphics[width=5.6in]{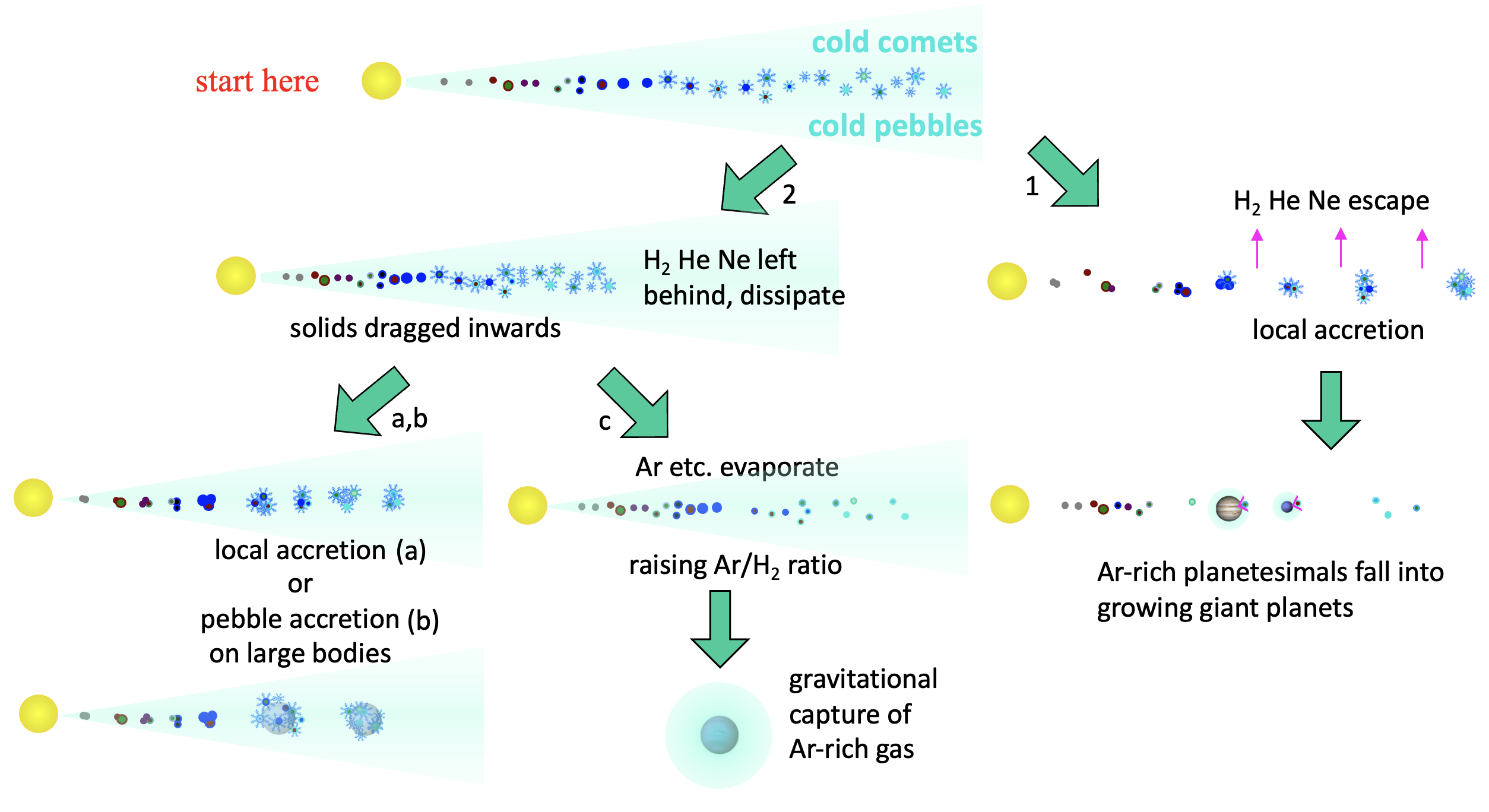} 
   \caption{Some possible pathways to concentrating super-volatile elements in gas giants. These are described in the text.
 In the figure, argon serves as a proxy for all the super-volatile elements. }
   \label{fig:two}
\end{figure}

As stressed above, the simplest way to separate C, N, S, Ar, Kr, and Xe from H$_2$ and He is to invoke temperatures
cold enough that all the elements other than H$_2$, He, and Ne freeze out quantitatively.
Once this is done, there are several plausible ways to separate the cold solids from the nebular gas.
Figure \ref{fig:two} illustrates some of these. 
\begin{enumerate}
\item H, He, and Ne can be photo-evaporated by EUV irradiation of the nebula \citep{Hollenbach2000,Guillot2006}. 
This increases the solid/gas ratio of the matter left behind.
A high solid/gas ratio promotes the aggregation of planetesimals by gravito-aerodynamic processes, such as the streaming instability \citep{Youdin2005} or turbulent concentration \citep{Estrada2016}.
The distant planetesimals could later be scattered and caught by the expansive envelopes of growing giant planets \citep{Pollack1996}, enriching their atmospheres. 
\item The icy pebbles or snowballs can spiral inward by aerodynamic drag \citep{Hayashi1985,Cuzzi2004}:
\begin{enumerate}
 \item This can increase the solid/gas ratio at some other place in the nebula closer to the Sun.
  Gravito-aerodynamic processes can then trigger planetesimal accumulation at the place they drifted to.
  \item The supervolatile-enriched snowballs can themselves be the pebbles in pebble accretion \citep{Ormel2017,Johansen2017} of a gas giant.
 \item The supervolatile-enriched snowballs could evaporate when they reach warmer climates, increasing the local Ar/He (e.g.) ratio in the nebular gas \citep{Cuzzi2004,Guillot2006}.
 The supervolatile-enriched nebular gas can then be captured by the gas giants when they acquire their envelopes.  
 \end{enumerate}
 \end{enumerate}

Jupiter's roughly uniform enhancement of the six super-volatile
elements, if it indeed extends to oxygen \citep{Li2020}, 
supports the widely-used working hypothesis that a global metallicity 
is useful for characterizing gas giants in general \citep{Thorngren2016}.  
Metallicity was originally defined by stellar astrophysicists as the sum of all elements other than H and He.
The only notable change when metallicity is applied to gas giants is that Ne is removed from the index of metals.
A Uranus atmospheric probe should aim to determine if the volatile elements are characterized by uniform metallicity.
It may long remain an open question whether uniform metallicity extends to the refractory elements \citep{Stevenson2020}.

\medskip

Neon does not condense at solar nebular temperatures.
Hence the null hypothesis is that Ne/He and $^{20}$Ne/$^{22}$Ne should be solar in Uranus's atmosphere.
Neon is much depleted in Jupiter's atmosphere because it follows helium raindrops into the abyss
\citep{Wilson2010}.  If we accept that equations of state are imperfectly understood, we must allow
for the possibility that neon could be depleted in the Uranian atmosphere.  
In principle neon can be separated by gravity from hydrogen and helium during hydrodynamic dissipation of the nebula,
and thereby enhanced in the residual gases captured by Uranus \citep[e.g.,][]{Mandt2020}.
In practice escape fluxes were likely much too big and gravity too weak to leave a measurable effect.   
Quantitative arguments are briefly discussed in Appendix \ref{Neon}.    

\subsection{Uranus inspired by Comet 67P/Churyumov-Gerasimenko}
\label{warm comets}

The only other well-characterized source of outer solar system noble gas abundances is Comet 67P/Churyumov-Gerasimenko \citep{Marty2017,Rubin2018,Rubin2019}.
Comet 67P/C-G was born in a warmer climate than the solids that evaporated in Jupiter's atmosphere.
In particular, Ar and Kr are significantly depleted in 67P/C-G, although much less so than they are in carbonaceous chondrites.
As we do not see sharp condensation temperatures preserved in moderately volatile elements in meteorites, 
we should not expect to find a single temperature describing partial condensation in a comet either. 
That said, the Ar and Kr in 67P/C-G appear to have been collected at temperatures warm enough that neither were native ices,
 although they may perhaps have guested in clathrates. 
 
Unlike Ar and Kr, Xe is present in 67P/C-G at a substantial fraction of its nominal solar abundance.
\citet{Rubin2018} report that the Xe/H$_2$O ratio in 67P/C-G's atmosphere was $2.4{\pm 1.1}\times 10^{-7}$.
For solar abundance, we follow \citet{Lodders2020}, who recommends O/Si and Xe/Si ratios of $16.6$ and $5.5\times 10^{-6}$, respectively.
At solar abundances, about 20\% of the oxygen is expected to be in refractory oxides.
Using 67P/C-G volatile abundances tabulated by \citet{Rubin2019}, 83\% of volatile oxygen is in H$_2$O.
If we allow another 5\% of the oxygen to reside in more refractory organic materials,
the expected solar Xe/H$_2$O ratio would be $5.5{\pm 2}\times 10^{-7}$.
Therefore, to first approximation, 67P/C-G is depleted in Xe with respect to the Sun by only a factor of two or three.
That a large fraction of the Xe originally in the gas condensed is an important constraint
when weighing the meaning of xenon's enormous isotope
anomalies, as these cannot be explained as an exotic trace phase, nor in any simple way as a highly fractionated remnant.
The isotopes will be discussed in detail in Section \ref{Xenon isotopes}.

If warm comets resembling 67P/C-G were more important volatile sources for Uranus than for Jupiter,
the different inheritances should be clearly visible in Ar, Kr, and Xe.  
 Figure \ref{fig:one} plots the predicted Ar, Kr, and Xe abundance patterns that would result if comets like 67P/C-G were
the major source of volatiles.
Two predictions are plotted.
The lower estimates (green squares) presume that the true bulk 67P/C-G C/O ratio is solar.
The higher estimates (black diamonds) scale volatile abundances to the 
subsolar C/O ratio observed in 67P/C-G's evaporated materials \citep{Mandt2020}.  
The predicted Xe abundance on Uranus is comparable to what it would be if the source were cold comets.
But the Kr/O and Kr/Xe ratios in 67P/C-G are distinctly subsolar,
and the Ar/O ratio is very subsolar; indeed, the predicted Ar/He ratio on Uranus would only slightly exceed the solar ratio, because 
 very little of Uranus's Ar would have been collected in solids. 
Such a noble gas pattern if seen might imply that Uranus was closer to the Sun than was Jupiter when Uranus captured its envelope,
or that accretion of Uranus predated envelope capture by Jupiter, or something else weirder still.

\section{Noble Clouds}
\label{Xenon Clouds}

\begin{figure}[htbp] 
   \centering
   \includegraphics[width=5in]{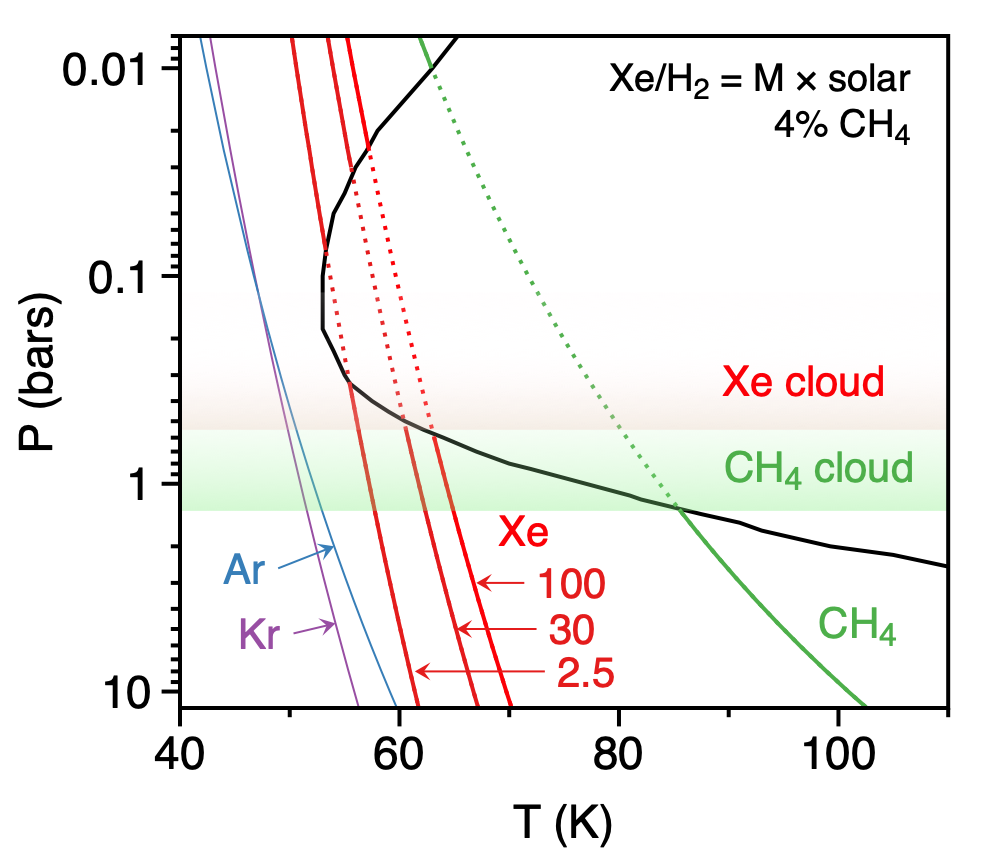} 
   \caption{Xenon clouds.  
 The plot compares condensation temperatures to a nominal $p\,$-$T$ profile of Uranus \citep{Sromovsky2011}.
 Condensation occurs when the ambient temperature is colder than the condensation temperature. 
 For methane we assume a 4\% volume mixing ratio below the cloud base.
 For Xe we show three different abundances:  $100\times$,  $30\times$, and $2.5\times$ solar.
 Xenon is expected to condense in all three cases.
 For Ar and Kr condensation we show only the highest abundance $100\times$ solar cases.
}
   \label{fig:three}
\end{figure}

Uranus is cold enough, and if any of our arguments for xenon's abundance hold,
xenon is abundant enough that it should condense,
a possibility that seems to have been overlooked \citep[e.g.,][]{Moses2020}.
Expected methane and xenon clouds are illustrated in Figure \ref{fig:three}.

In Figure \ref{fig:three} we use a $p$-$T$ profile from \citet{Sromovsky2011}, which for $p<1$ bar 
is essentially identical to the $p$-$T$ profile found by \citet{Lindal1987}.
For the vapor pressures of noble gases we fit a simple Antoine equation
\begin{equation}
\label{equation:one}
\log_{10}P_{\mathrm{sat}} = A - {B\over T+C}
\end{equation}
to sublimation vapor pressure data plotted by \citet{Ferreira2008}.
Our Antoine parameters for Xe, Ar, Kr, and CH$_4$ are listed in Table~1.
For CH$_4$ we fit Antoine parameters for the vapor pressure over the solid 
that were originally published by \citet{Karwat1924}.
There must be data somewhere that go to lower $T$, but our fit is adequate for our purposes.

\begin{table}[htp]
\caption{Antoine parameters used here}
\hskip-1.6cm   
\begin{tabular}{l r r r r r r}
\hline
Species  & A~~ & B~~ & C~~ & $T_{min}$ & $T_{max}$ & Reference\\
CH$_4$  & 4.18 & 439 & -5 & 77 & 110 &  Dortmund Data Bank$^a$ \\
Ar  & 4.63 & 380 & -4 & 37 & 85 & \citep{Ferreira2008} \\
Kr  & 5.735 & 671.3 & 4 & 37 & 90 & \citep{Ferreira2008}\\
Xe  & 3.74 & 633 & -8 & 50 & 90 & \citep{Ferreira2008} \\
\hline
\multicolumn{7}{l}{$a$ -- DDB's reference for vapor pressure over methane ice is \citet{Karwat1924}.}\\
 \end{tabular}
\label{table_one}
\end{table}

Condensation occurs when the saturation temperature is higher than the ambient temperature.
The saturation temperature of a gas for a given partial pressure $P$ 
is obtained by inverting Eq \ref{equation:one}
\begin{equation}
\label{equation:two}
T_{\mathrm{sat}} = {B\over A - \log_{10}P} - C.
\end{equation}
Saturation temperatures for Ar, Kr, Xe, and CH$_4$ are plotted against the nominal Uranian $p$-$T$ profile in Figure \ref{fig:three}.
For CH$_4$ condensation we assume a 4\% mixing ratio below the clouds.  
For Ar and Kr we assume a high abundance of 100$\times$ solar in order to maximize their chances of condensing.
For Xe we plot enhancements of 100, 30, and 2.5 times solar.  
These correspond to volume mixing ratios of 31, 9, and 0.8 ppbv, respectively. 
For any plausible abundance, Xe is expected to condense as an ice.
For 10 ppbv ($33\times$ solar), Xe condenses at $T\sim 61$ K at altitudes above $\sim 0.5$ bars.
Because it condenses, there should be no expectation that Xe is uniformly mixed below the clouds.
Given xenon's very low abundance, it seems likely that condensation is on methane ice grains or other
available particles rather than as pure clouds.

\subsection{Xe fractionation in condensation}

Condensation can cause fractionation. %
At 60 K, the difference in vapor pressures between $^{130}$Xe and $^{136}$Xe is expected to be of the order of $1\permil$,
using data and equations given by \citet{Alamre2020}.
At peak condensation, only 2\% of Xe remains in the vapor phase for a nominal $40\times$ solar enhancement, which
implies that the vapor's $^{136}$Xe/$^{130}$Xe ratio could be enriched $\sim\! \left(1/0.02\right)^{0.001}\sim\!4\permil$ by Rayleigh distillation.  
This is very small compared to the typical magnitude of Xe's isotopic fractionations (see below) or the plausible capabilities of any
Uranus Probe instrument.
We conclude that isotopic fractionation during condensation of Xe is unlikely to be important.

\subsection{Ar and Kr}
As shown in Figure \ref{fig:three}, neither argon nor krypton condense in the nominal atmosphere, even if we assume high end 
enhancements of $100\times$ solar. 
This makes them the most useful elements for assessing the role of cold comets.
If Uranus accreted its envelope of highly volatile elements from the same reservoir from which Jupiter accreted its envelope,
the Ar/Kr ratio would be solar and the Ar/H$_2$ ratio would be enhanced over solar by a factor of order $30-100$.  
The expected abundance pattern is shown by the red open circles in Figure \ref{fig:one}.
If instead Uranus accreted its noble gases as warm comets resembling 67P/C-G, the Ar/Kr and Kr/Xe ratios would
be subsolar and the Ar/H$_2$ ratio would be solar.  
This abundance pattern is shown by the open green and black symbols in Figure \ref{fig:one}.

It may be possible that the Uranian winter poles get cold enough for Ar and Kr to condense.
Figure \ref{fig:three} shows that condensation near the tropopause becomes possible if Ar and Kr abundances are at the
high end of our predictions ($\sim\!500$ ppmv Ar) and ambient temperatures fall below $\sim\!48$ K. 
If Ar and Kr do condense, they too might not be globally mixed.
Also, if Ar does condense, it must be abundant, and therefore its clouds should be relatively substantial.
Polar argon clouds might be detectable remotely as spring approaches, or from an orbiter via stellar occultation. 
   
\section{Xenon isotopes}
\label{Xenon isotopes}

\begin{figure}[!htb]
 \centering
 \begin{minipage}[c]{0.56\textwidth}
   \centering
\includegraphics[width=1\textwidth]{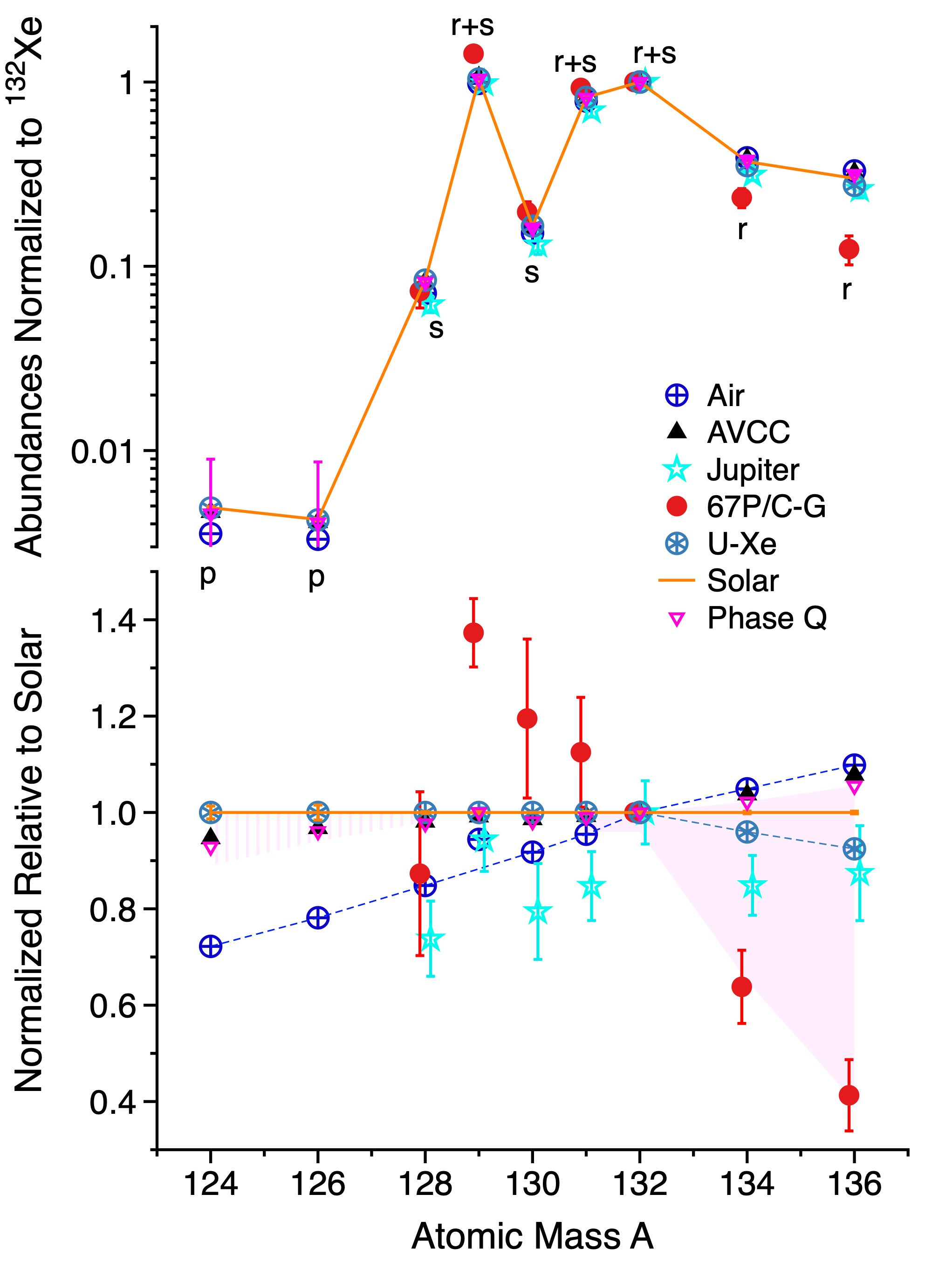} 
 \end{minipage}
 \begin{minipage}[c]{0.42\textwidth}
   \centering
  \caption{Selected xenon isotopes in the Solar System.
   In the top panel abundances are normalized to $^{132}$Xe.
   This shows which nuclei are favored by nucleosynthesis.  
    Nucleosynthetic sources of each isotope are indicated: ``r'' for rapid neutron addition and ``s'' for slow, and ``p'' for rapid proton addition. 
   U-Xe is the reconstructed birth composition of Xe in air \citep{Pepin2000}.
   Phase Q and AVCC-Xe are representative of carbonaceous chondrites \citep{Pepin1991,Busemann2000}.
   The bottom panel renormalizes the top panel to solar abundances, which facilitates comparing patterns.
  The Xe in comet 67P/C-G stands out as being rich in the partially radiogenic $^{129}$Xe 
  ($^{129}$I, 15.7 Myr half-life, also r+s)
    and being extremely deficient in the heavy r-only isotopes $^{134}$Xe and $^{136}$Xe \citep{Marty2017}.
   Pink shading suggests the range of what might be found on Uranus.
    }
 \end{minipage}
\label{fig:four}
\end{figure}

Xenon's nine stable isotopes vary enormously in the Solar System and often display very large signals, which
make them excellent targets for a Uranus atmosphere probe.
Figure \ref{fig:four} plots selected Xe isotopic data.
The upper panel plots abundances of the different isotopes normalized to $^{132}$Xe,
while the lower panel renormalizes the data to solar abundances, which facilitates comparing patterns.  
   The selection includes the Sun \citep[through the solar wind,][]{Meshik2020};
   average carbonaceous chondrites \citep[AVCC,][]{Pepin1991}; 
   the ubiquitous Phase Q \citep{Busemann2000}; Jupiter as measured by the Galileo Probe \citep{Mahaffy2000};
  comet 67P/C-G \citep{Marty2017}; Earth's air; and U-Xe \citep[air's primary Xe,][]{Pepin2000}. 
These particular xenons were selected to tell a story.
   
Meteorites are diverse and the noble gases found in them even more diverse \citep{Gilmour2010,Marty2022}.
Here we opt for a greatly oversimplified discussion.
Much of the Ar, Kr, and Xe in chondrites is found in a refractory organic material called Phase Q \citep{Busemann2000}.
Compared to the solar wind, all 3 heavy noble gases (Ar, Kr, Xe) are isotopically mass fractionated in Phase Q by about $\sim\!1\%$ per amu,
with heavier isotopes preferred and heavier elements very strongly preferred.
The mass fractionation can be ascribed to many possible causes.
Three of note are ion chemistry during synthesis of organic matter \citep{Frick1979,Kuga2015,Kuga2017}; gravitational settling within porous planetesimals \citep{Ozima1980,Zahnle1990}; and fractionation during escape, probably from the planetesimal itself \citep{Benedikt2020}.
 
 Average carbonaceous chondritic Xe (AVCC Xe)
 is distinctly less mass fractionated ($\sim\!0.6\%$ per amu) than Phase Q, 
 but it is also enriched in Xe's two heaviest isotopes \citep{Gilmour2010,Mandt2020}.
The excess $^{134}$Xe and $^{136}$Xe is called Xe-H. 
 When allowance is made for mass fractionation, AVCC Xe is also seen to be enriched in the light isotopes $^{124}$Xe and $^{126}$Xe.
  The excess of $^{124}$Xe and $^{126}$Xe is known as Xe-L. 
 When the H and L excesses occur together, as they usually do in meteorites, the excesses are known as Xe-HL.
 Xe-HL is carried in presolar grains, chiefly nanodiamonds, silicon carbides, and graphite \citep{Huss2008,Ott2012,Mandt2020}.
  Many meteorites loosely resemble AVCCs in these ways \citep{Huss1996},
 although the mass fractionations in enstatite chondrites and ureilites are smaller than in AVCCs,
 and some enstatite meteorites (e.g., the aubrite Pesyanoe) carry solar Xe.
   
  Atmospheric Xe is very strongly mass fractionated and polluted with radiogenic debris, but 
  its real strangeness lies in its being deficient in the 
  heavy isotopes $^{134}$Xe and $^{136}$Xe compared to the Sun. 
   Atmospheric Xe evolved by addition of spontaneous fission of $^{244}$Pu (80 Myr half-life)
   and by preferential escape of light isotopes over the course of 2 billion years \citep{Avice2018,Zahnle2019}.
   One reconstructs Earth's original Xe---called ``U-Xe''---by reversing this history \citep{Pepin1991,Pepin2000}.  
   How this is done is sketched out in Appendix \ref{U-Xenon}. 
   It is notable that U-Xe does not appear to have been deficient in the light isotopes $^{124}$Xe and $^{126}$Xe;
   i.e., Earth is deficient in Xe-H but not in Xe-L.
   U-Xe had been very hard to understand:
     it seemed impossible to subtract Xe-H from Earth to turn solar Xe into U-Xe, and it is impossible to add a small sprinkling of exotic stardust to a Sun-sized reservoir of U-Xe to change it into solar Xe.

   Results from the Rosetta mission to Comet 67P/C-G,
  when taken at face value, 
    provide the key to unlocking U-Xe \citep{Marty2017,Rubin2018}.
   The xenon evaporating off comet 67P/C-G qualitatively resembles U-Xe in its deficiency of $^{134}$Xe and $^{136}$Xe,
   but the magnitudes of the isotope anomalies in 67P/C-G are truly gigantic (Figure \ref{fig:four}).
 If one neglects $^{128}$Xe (which is relatively rare and suffers from confusion with S$_4$),
  the reported data could be fit to an extreme negative mass fractionation of the order of $-14\%$ per amu.
  Malfunction seems unlikely: the instrument also returned a wholly credible measurement of an on-board Xe calibration sample \citep{Marty2017}.
   Fractionation also seems unlikely:  the nearly solar Xe/H$_2$O ratio in 67P/C-G's vapors 
   implies that we are looking at a large fraction of the comet's original xenon,
   rather than at a highly fractionated tail of something that had once been much more abundant. 
   Nonetheless, one might be tempted to conjure an {\em ad hoc} mechanism for grossly fractionating Xe
   were it not for the fact that 67P/C-G's xenon resolves the mystery of U-Xe.
   
    If we accept the Rosetta data, Xe in 67P/C-G is either extremely deficient in $^{134}$Xe and $^{136}$Xe or it is 
   extremely enriched in $^{129}\mathrm{Xe} - ^{132}\!\mathrm{Xe}$.
  For convenience, we will abbreviate the strange Xe of 67P/C-G as ``CG-Xe''
   (most single letters of the alphabet are already taken). 
   Within the generous uncertainties,
   it is possible to create U-Xe by mixing $\sim\!22\%$ CG-Xe with chondritic Xe \citep{Marty2017}.
   In other words, it now appears that a tributary to U-Xe does exist as a pure component in the Solar System.
   U-Xe itself now appears to represent a pooled average of a wider span of primary compositions, and as such may be unique to Earth.

\subsection{Nucleosynthetic xenon}

  The differences between 67P/C-G and the solar nebula are almost certainly nucleosynthetic. 
  Nucleosynthesis in elements heavier than iron is categorized by process.
   The p-process refers to rapid proton addition, the r-process to rapid neutron addition, and
   the s-process to slow neutron addition \citep{Arnould2007,Kappeler2011}.
   Xenon is usually classified as an r-process element \citep{Arnould2007},
   but because it has nine stable isotopes, the full story is more complicated: 
   there are at least two r-processes, many s-processes, and a p-process \citep{Gilmour2007,Marty2017,Avice2020}. 
   Figure \ref{fig:four} indicates the processional source or sources of each of the isotopes. 
   
   The main r-process 
   creates a prominent abundance peak centered on $^{128}$Te, $^{129}$I (which eventually decays to $^{129}$Xe, $\tau_{1/2}=15.7$ Myrs), $^{130}$Te, and $^{131}$Xe
    \citep{Arnould2007}.
   It is responsible for $^{129}$Xe, $^{131}$Xe, and much of $^{132}$Xe.
    The neutron-rich isotopes $^{134}$Xe and $^{136}$Xe (Xe-H) are made at least in part by a different r-process because,
     as clearly seen in Figure \ref{fig:four}, the abundances of $^{134}$Xe and $^{136}$Xe scale independently of the main r-process.
   \citet{Gilmour2007} and \citet{Avice2020} call this other r-process the ``h-process'' to reflect its special role in forging Xe-H. 
  
   The s-process is usually held uniquely responsible for $^{128}$Xe and $^{130}$Xe because direct 
  r-process formation is blocked by stable $^{128}$Te and $^{130}$Te, respectively.
   However, the s-process is contingent on the starting composition \citep{Kappeler2011}.
   For example, neutron capture by $^{129}$Xe creates $^{130}$Xe.
   Thus the relatively high abundance of $^{130}$Xe in 67P/C-G seems to imply that the s-process acted on a glut of main r-process isotopes.
   When this observation is reconciled with the depletion of $^{134}$Xe and $^{136}$Xe, 
   it seems reasonable to describe CG-Xe as being s-process enriched \citep{Marty2017,Rubin2018}.
    
   The light isotopes $^{124}$Xe and $^{126}$Xe (Xe-L), which are very rare, must be made by a p-process.
  Correlated excesses of heavy and light Xe isotopes are prominent in presolar diamonds \citep{Ott2012}.
   For our purposes, the H-L correlation is something to look for on Uranus.
   Hence we have sketched in pink a possible p-process deficit on Uranus in Figure \ref{fig:four}.
  Because the p-process nuclei are rare, they may set a performance floor on the Probe.

 \subsection{Strange xenon in the solar system and beyond}
   
\begin{figure}[!htb]
 \centering
 \begin{minipage}[c]{0.54\textwidth}
   \centering
  \includegraphics[width=1\textwidth]{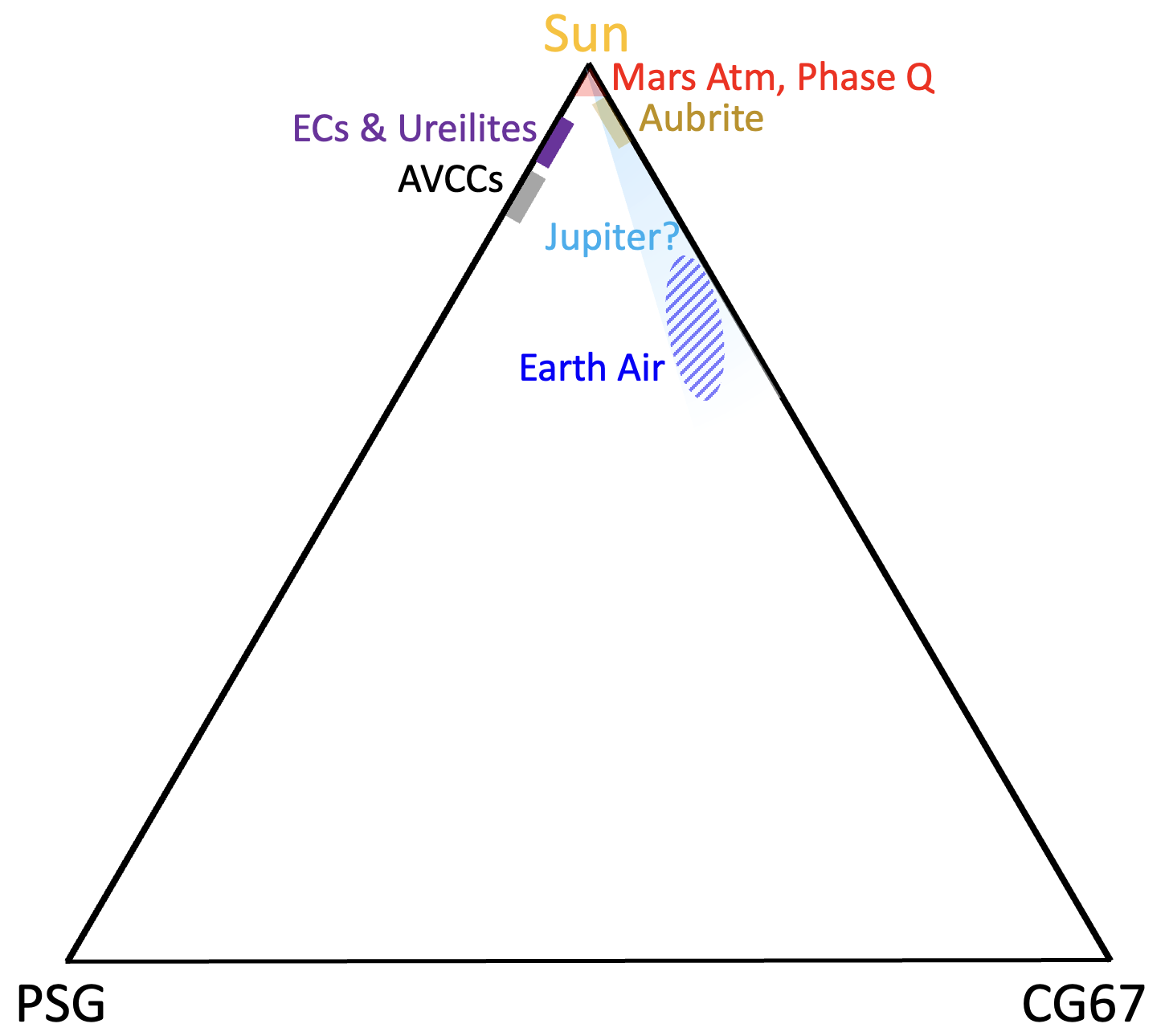} 
 \end{minipage}
 \begin{minipage}[c]{0.44\textwidth}
   \centering
  \caption{A schematic ternary plot of Xe in the solar system.  
 The three vertices---solar, presolar grains (PSG), and comet P67/C-G---
  represent isotopic anomalies that can only be explained by distinct nucleosynthetic reservoirs.
  Earth's birth Xe (U-Xe) is of mixed parentage.
  Data for enstatite chondrite (EC, S.\ Oman), aubrite (Pesyanoe), and AVCC are from \citet{Pepin1991};
  the ureilite and Phase Q are from \citet{Busemann2000};  Mars \citep{Peron2022}.
  All reservoirs other than P67/C-G have been corrected for evident mass fractionation.
Jupiter is extremely uncertain but intriguing.  
  }
\label{fig:five}
 \end{minipage}
\end{figure}

   The simplest classification scheme is a ternary plot with the Sun, presolar grains (PSG), and 67P/C-G as vertices
   (Figure \ref{fig:five}). 
   Chondrites appear to have captured nebular xenon resembling Phase Q in organic matter.
   This xenon was mass fractionated to varying degrees by processes probably linked to the capture process itself.
   Presolar grains added excess Xe-HL to chondrites.
   The third distinct reservoir seen in 67P/C-G and Earth's air is labeled ``CG67'' on Figure \ref{fig:five}.
    Earth's original Xe (U-Xe) can be described as a solar base with addition of $\sim\!14\%$ 67P/C-G-like xenon,
   or as an AVCC-like base with addition of $\sim\!22\%$ 67P/C-G-like xenon \citep{Marty2017}. 
   
   A relationship, if any, between the end-member reservoirs seen in Xe and the more familiar NC/CC isotopic dichotomy
    seen in refractory inner solar system materials is unclear.  The general trend may be the same, as s-process nuclei seem more
    prominent nearer the Sun, and both trends could be consistent with a late infall of s-process-rich material on the inner solar system
    \citep{Nanne2019}.
    However, the amplitude of the trend in Xe is orders of magnitude larger.
  The different xenons might characterize nebular reservoirs on larger spatial or temporal scales than the NC/CC dichotomy. 

\medskip

      The nucleosynthetic anomalies seen in 67P/C-G's xenon are so large that they almost seem to have come from another solar system.
     One way to account for the strange composition of 67P/C-G would be if it had been scattered out of, or captured from,  
      another solar system in the Sun's birth cluster.
    This other solar system would by construction have precipitated out of a different local mix of stellar effluents.
 Another possibility would be if the Sun accreted a fresh cloud of gas after the original solar nebula had dissipated,
      sparking a second burst of planetesimal formation within the solar system starting from a different base composition.
 A third option inverts the chronology: perhaps CG67 condensed extremely early, {\em before} the solar nebula was enriched by adding a fresh injection of 
    Xe-H-rich gas.
         
   The origin of the different xenons remains a mystery.
   AGB stars are known sites of s-process nucleosynthesis \citep{Kappeler2011}.
   Strange xenon in 67P/C-G could record effluent from an AGB star that had no chance 
   to mix with the solar nebula's other materials.
  There are planetary nebulae observed to be enriched in Xe by an order of magnitude,
 which is ascribed to s-process nucleosynthesis in the particular AGB star that is now shedding its atmosphere 
 \citep{Sharpee2007}.
    It is also plausible that 67P/C-G Xe was relatively young.  Youth is suggested by the anomalously high abundance of $^{129}$Xe in CG-Xe, which 
  can be explained if 67P/C-G captured live $^{129}$I more efficiently than it captured Xe \citep{Marty2017,Avice2020}.
     
  As \citet{Marty2017} emphasize, it takes very little CG-Xe to make a big mark on Earth's Xe,
   because Xe is very abundant in 67P/C-G.
 Hence the signature of 67P/C-G-like material in Earth is expected to be very small in more refractory elements, including water.
 Nonetheless, what is valid for Xe will be valid to some extent for other elements.
  Subtle terrestrial isotopic anomalies, such as those seen in ancient cratonic ruthenium \citep{Fischer-Godde2020},
  could stem from the same s-process-rich source as Earth's strange xenon.
  Non-solar abundances might not be limited to heavy elements.
  Carbon in particular is linked to s-process synthesis in AGB stars because the neutrons come from $^{13}$C \citep{Kappeler2011},
  which hints that the C/O ratio may not have been the same everywhere in the solar system.
  It is possible that Uranus's composition was affected. 
  
\subsection{Krypton}

\begin{figure}[!htb]
 \centering
 \begin{minipage}[c]{0.58\textwidth}
   \centering
\includegraphics[width=1\textwidth]{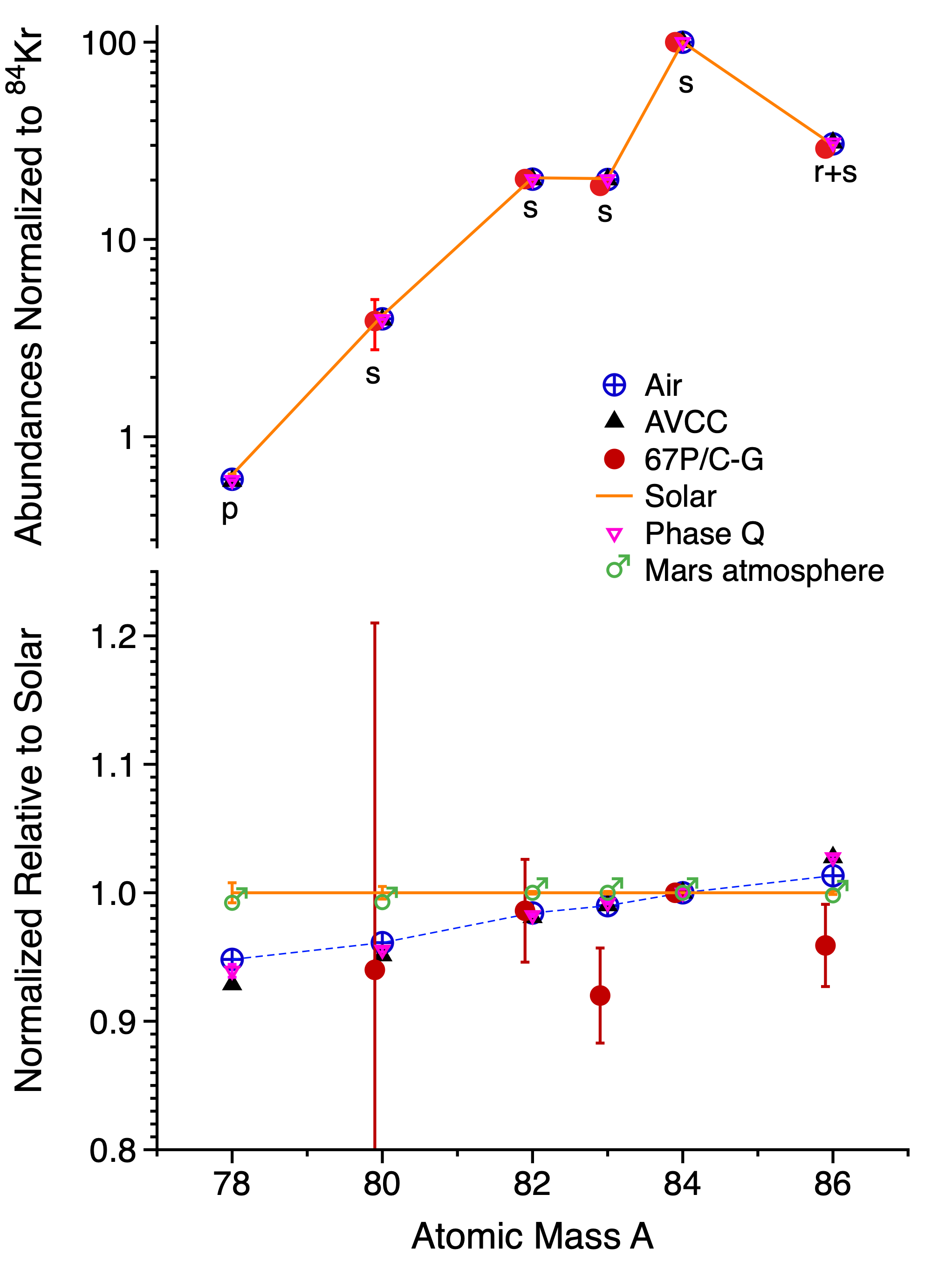} 
 \end{minipage}
 \begin{minipage}[c]{0.40\textwidth}
   \centering
  \caption{Krypton isotopes in the Solar System: abundances and nucleosynthetic sources.
   The top panel normalizes abundances to $^{84}$Xe.
   This shows which nuclei are favored by nucleosynthesis.  
   The bottom panel is normalized to solar abundances, which facilitates comparing patterns.
   Phase Q, AVCC-Xe \citep{Pepin1991,Busemann2000}, and air are similar---all are mass-fractionated with respect to the Sun,
   although air slightly less so.
   Mars's atmosphere is solar and its interior appears chondritic \citep{Peron2022}.
   67P/C-G appears to show distinct nucleosynthetic anomalies \citep{Rubin2018}.
    }
\label{fig:six}
 \end{minipage}
\end{figure}

Krypton's isotopic patterns are muted compared to xenon's (Figure \ref{fig:six}).
As noted above, Kr's isotopes, like Xe's, are mass fractionated by as much as 1\% per amu in Phase Q and many chondrites.
 Krypton also shows some small isotopic anomalies, chiefly in $^{83}$Kr and $^{86}$Kr.
 These show as deficits in 67P/C-G.
 Earth's mantle also shows a small deficit in the neutron-rich isotope $^{86}$Kr \citep{Peron2021}, which parallels what
 is seen in xenon. 
 Small $^{83}$Kr {\em excesses} are sometimes seen in interstellar grains \citep{Huss2008} that are enriched in Xe-H,
 a mirror image pattern that also parallels what is seen in xenon. 
Krypton is made by multiple s-processes (both main and ``weak,'' each with several options for branching), an r-process, and a p-process \citep{Kappeler2011}. 
It is therefore not surprising to see different nucleosynthetic patterns in different materials.
  \citet{Rubin2018} and \citet{Avice2020} devise detailed mixing models involving both Kr and Xe and multiple nucleosynthetic sources.
In these models, 80\% of the Xe in 67P/C-G is novel but only 5\% of the Kr is. 
If so, it is plausible that the gas 67P/C-G accreted from had an elevated Xe/O ratio, which supports the
hypothesis that $^{129}$I was alive when 67P/C-G accreted. 
 
  \subsection{Jupiter, again}
  
 There is a slight hint of $^{134}$Xe and $^{136}$Xe deficiency in Jupiter, seen in Figure \ref{fig:four}.
  Although the data aren't precise enough to warrant any conclusions, one could easily imagine that one is
  seeing a $10\%$ contribution from 67P/C-G-like Xe, 
  which is where we plot Jupiter on Figure \ref{fig:five}.
  Obtaining a clear picture of Xe isotopes on Uranus could go a long way toward sorting out 
  the role of 67P/C-G-like and the importance of alternative cosmic abundances in the Solar System.

\section{Conclusions}
\label{Conclusions}

Noble gas elemental abundances and Xe isotopes have much to say about the cosmic provenance of Uranus
if we take the opportunity to ask. 
Argon and krypton elemental abundances
will reveal the extent to which Jupiter's envelope's apparently uniform abundances of C, N, S, Ar, Kr, and Xe translate to
an ice giant, and they provide a solar system test of the astrophysically convenient metallicity hypothesis \citep[c.f.,][]{Thorngren2016}.
 Neon, if it should prove overabundant, would provide a window into the mechanism by which Uranus attracted its nebular envelope.

Xenon condensation is an anticipated feature of Uranus.
As seen with ammonia on Jupiter, the true abundance of a condensible gas in a gas giant is likely to be variable and hence difficult to determine with any confidence \citep{Guillot2020}. 
 A single probe may give a misleading picture of xenon's abundance, even for measurements taken below the clouds.  
Instead, xenon becomes a tracer of cloud dynamics and atmospheric circulation,
which is not a role obviously well suited to a species that in practice cannot be mapped. 
 On the other hand, condensation concentrates Xe much more effectively than a noble gas enrichment cell.
 The clouds present an opportunity for precisely measuring Xe's isotopes that a suitably designed probe could exploit;
 in effect, by doing intentionally what the Pioneer Venus Probe did accidentally for D/H \citep{Donahue1982}.
 
What can be learned from Xe's isotopes? 
To first approximation, xenon's isotopes appear to track three distinct reservoirs of solar nebular gas.
One resembles the Sun and, when mass fractionated, many chondrites.
A second, found in presolar grains, tracks a particular style of nucleosynthesis that favors
the most proton-rich and most neutron-rich isotopes, and then binds them in refractory carbon-rich grains. 
The second source exists only in trace amounts in the inner solar system and is not expected to be measurable on Uranus.  
The third source is present at nearly solar abundance in 67P/C-G, but is also discoverable in Earth's air.
Xenon's high abundance in 67P/C-G suggests that, like the solar nebula, the third source was condensed from a gas.  
It is an intriguing observation that the third source is highly depleted in Xe's heavy r-process-only isotopes (Xe-H) in approximate mirror image
to how those same isotopes appear enriched in the presolar grains.
The mirror image relationship does not extend to the proton-rich nuclei.
There is an underlying story behind this pattern that has yet to be told.

When we focus in on 67P/C-G as a tracer of the third source, we see hints of new matter being added at a relatively late time
in the solar nebula's evolution.
The very high $^{129}$Xe in 67P/C-G can be explained if iodine condensed more efficiently than xenon ---
in itself a reasonable request --- provided that the gas the comet condensed from was young enough that 
the newly synthesized $^{129}$I (15.7 Myr half-life) had yet to decay.
In other words, the gas was young when 67P/C-G was made.
Whether comets like 67P/C-G condensed within our solar system in a late burst of planetesimal accretion,
or whether they fell fully formed from some other solar system, are open questions.

 It is to be determined if the third source contributed significantly to Jupiter and Uranus. 
Xenon's isotopic anomalies in the Solar System are big enough to suggest that something interesting ought to be detectable in Uranus, and when found, 
 will reveal how the nucleosynthetic sources of Uranus's materials resemble or differ from those of the inner solar system, Jupiter,
  or even the Sun itself.

\medskip
{\it Acknowledgments.}
We thank two anonymous referees for their thoughtful and substantive reviews, as well as for their advice on
what it should be called. 
This research did not receive any specific grant from funding agencies in the public, commercial, or not-for-profit sectors.
\newpage

\appendix
\restartappendixnumbering

\section{On Neon}
\label{Neon}

It has been suggested that noble gases could have been mass fractionated during the  
photoevaporative hydrodynamic escape of hydrogen and helium from the nebula \citep{Mandt2020}.
Because Ne is heavier than H$_2$ and He, it can fall behind during dissipation of the nebula,
which would make Ne in the remnant of the nebula more abundant and isotopically heavier.
Neon in Uranus's atmosphere might keep a record of these events.
However, diffusive separation is difficult because the gravitational force responsible for diffusive separation would be weak \citep{Guillot2006}.
To quantify the discussion,
it is helpful to construct an example of diffusive separation of Ne in an escaping hydrodynamic wind of nebular gases.

The cross-over mass $m_c$ is defined as the heaviest isotope that can escape in a hydrodynamic wind.
If $m<m_c$, the efficiency of escape is proportional to $m_c-m$.
By definition, there is no escape if $m>m_c$.
The cross-over mass for neon can be approximated by \citep{Hunten1987}
\begin{equation}
\label{crossover mass}
m_{\mathrm{c}} = m_{\mathrm{n}} + {\phi k_{\rm B} T N_{\mathrm{Avo}}\over g b_{\mathrm{n,Ne}}  } ,
\end{equation}
where $m_{\mathrm{n}}\approx 2.4$ amu is the mean molecular weight of the nebular gas and $m_{\mathrm{Ne}}=20$;
 $b_{\mathrm{n,Ne}} \approx 4.3\times 10^{17}\,T^{0.74}$ cm$^{-1}$s$^{-1}$ is the binary diffusion coefficient between Ne and the nebular gas \citep[it is nearly the same for Ne and either H$_2$ or He,][]{Marrero1972};
and $N_{\mathrm{Avo}} = 6.022\times 10^{23}$ is Avogadro's number.
The fractionation factor, used in Rayleigh distillation, is the ratio of escape probabilities of two species of different mass, 
\begin{equation}
\label{fractionation factor}
x_{\mathrm{n,Ne}} = { m_{\mathrm{Ne}} - m_{\mathrm{n}} \over m_{\mathrm{c}} - m_{\mathrm{n}} } \geq 0.
\end{equation}
Gravity is approximated by $g = G M_{\odot} z/R^3$, 
where $M_{\odot}$ is the mass of the Sun, $R$ is the radial distance from the Sun, and $z$ is the height above the disk plane.
To complete the illustration, take $R=10$ AU and $z/R=0.2$, for which $g \approx 10^{-3}$ cm s$^{-2}$.
The escaping flux $\phi_{\mathrm{n}}$ of H$_2$ and He can be estimated from the total amount of material that needs to escape divided by the area from which it escapes and the time taken to escape.
We might expect escape of a Jupiter's mass of material from an annulus of width $\Delta R=0.2R$ over 5 million years.
With these choices, 
\begin{equation}
\label{phi}
 \phi_{\mathrm{n}} = 1.1\times 10^{11} \left( M \over M_{\rm Jup} \right) \left( 5\mathrm{~Myrs} \over t \right) \left( 20 \mathrm{~A\!U}^2 \over R\,\Delta R \right) \,\, \mathrm{cm}^{-2}\mathrm{s}^{-1}  .
\end{equation}
The crossover mass would then be $m_c \approx 4.4\times 10^4\left(T/30\right)^{0.26}$, which is so high that very few neon atoms would be left behind.  The fractionation factor between Ne and the hydrogen/helium is $x_{\mathrm{n,Ne}}\approx 0.0004$.  
If Uranus captured the last 0.3\% of a Jupiter's mass of gas,
we can use Rayleigh distillation to estimate that the Ne/H$_2$ ratio would be raised by 
a factor $300^{0.0004}=1.0023$, which is quite negligible.

Nor is neon expected to fractionate on infall: accretion of an Earth mass of nebular gas over 3 million years on Uranus corresponds
to an influx $\phi_{\mathrm{n}} \sim -2\times 10^{17}$ cm$^{-2}$s$^{-1}$, which for $g=1000$ cm s$^{-2}$ implies a crossover mass of order $10^5$ amu.  
To achieve significant enhancements of Ne/H$_2$ or $^{22}$Ne/$^{20}$Ne by diffusive separation of gases 
 requires much more modest flows acted on by a strong gravitational field, as might be possible
 from a slowly dissipating, relatively low mass heliocentric torus not too distant from the Sun.
 As such a scenario is not wholly unimaginable, neon may hold surprises.

\section{On U Xenon}
\label{U-Xenon}
\restartappendixnumbering

Earth is markedly deficient in xenon's r-process-only heavy isotopes $^{134}$Xe and $^{136}$Xe,
although this fact is hidden under the mass-dependent fractionation that also characterizes atmospheric Xe.
By construction, U-Xe is designed to recover the composition of Xe in air before radiogenic isotopes
 were added and before atmospheric xenon was acted upon by fractionating escape.
U-Xe was originally conceived as the true solar Xe, then later as a kind of Xe that existed in pure form 
somewhere in the Solar System \citep{Pepin1991,Pepin2000,Pepin2006}.  
It has not yet been found.
What happened instead was that Rosetta revealed in 67P/C-G a world in which
the depletions of $^{134}$Xe and $^{136}$Xe far exceed those of U-Xe.
Evidently U-Xe was not a pure component, but rather a mixture of older components, at least one of which resembled 67P/C-G in being very
deficient in $^{134}$Xe and $^{136}$Xe \citep{Marty2017,Rubin2018}.   
It now seems that U-Xe marked a stage in Earth's early atmospheric evolution, rather than its beginning.

U-Xe is reconstructed from atmospheric xenon by
 (i) undoing the mass-dependent fractionation and 
 (ii) by subtracting radiogenic Xe.
The chief radioactive sources, $^{129}$I ($\tau_{1/2}=15.7$ Myr) and $^{244}$Pu ($\tau_{1/2}=80$ Myr), are both short-lived.
It is now known that half of xenon's mass fractionation took place in the Archean between 3.5 Ga and 2.4 Ga
\citep{Avice2018,Ardoin2022}, and it is reasonable to infer that much of the rest of the fractionating escape took place 
under similar conditions between 4.4 Ga and 3.5 Ga. 

Isolating air's small $^{129}$Xe excess is straightforward, but thanks to the large $^{129}$Xe excess in 67P/C-G, the classic interpretation 
of excess $^{129}$Xe in the atmosphere as $^{129}$I decay {\it in situ} on Earth \citep{Ozima2002} is no longer on firm footing \citep{Marty2017}.
Isolating air's fissiogenic xenon is more challenging because fission spawns a broad spectrum of isotopes that have to be separated from mass fractionation. 
 $^{244}$Pu is predominantly an $\alpha$-emitter.
 The branching ratio for spontaneous fission is $1.23\times 10^{-3}$ \citep{Ozima2002}.  
The Xe isotope most affected by fission is $^{136}$Xe.
It is produced in 5.6\% of the spontaneous fissions \citep{Ozima2002}.
It is convenient to scale fission products in relation to $^{136}$Xe.
Plutonium's daughters are listed in Table \ref{table_nine}.

\begin{table}[htp]
\caption{The Nine Isotopes of Xenon}
\footnotesize
\flushleft
\begin{tabular}{lllll  |  ll}
\hline
$j$ & Air$^a$  & SW$^b$ & AVCC$^c$ & U-Xe$^a$ & Pu-Xe$^d$ & Xe-H$^e$  \\
 124 & $2.337\pm 0.007$ & $2.97\pm 0.04$ & $2.851\pm 0.051$ & $2.928 \pm 0.01$  & 0 & 0   \\
 126 & $2.18\pm 0.011$  & $2.56\pm 0.04$ & $2.512\pm 0.04$ & $2.534 \pm 0.013$ & 0 & 0   \\
 128 & $47.15\pm 0.047$ & $51.0\pm 0.1$ & $50.73\pm 0.38$ & $50.83 \pm 0.06$ & 0 & 0   \\
 129 & $649.6\pm 0.58$  & $631.4\pm 1.3$ & $653\pm 17$ & $628.6 \pm 0.6$  & $0.048\pm 0.055$ & $0.196\pm 0.02$ \\
 130 & $100\phantom{\pm 0.1}$  & $100 \pm 0.34$ & $100$  & $100$  & 0 & 0  \\
  131 & $521.3\pm 0.59$ & $501.0\pm 1.2$ & $504.3\pm 2.8$ & $499.6 \pm 0.6$  & $0.248\pm 0.015$ & $0.171\pm 0.015$ \\
 132 & $660.7\pm 0.53$  & $606.3 \pm 1$ & $615\pm 2.7$  & $604.7 \pm 0.6$  &  $0.893\pm 0.013$ & $0.154\pm 0.01$  \\
 134 & $256.3\pm 0.37$  & $224.2 \pm 0.6$ & $235.9\pm 1.3$ & $212.6 \pm 0.4$  & $0.930\pm 0.005$ & $0.694\pm 0.01$   \\
 136 & $217.6\pm 0.22$  & $181.8 \pm 0.5$ & $198.8\pm 1.2$  & $165.7 \pm 0.3$ & 1 & 1   \\
 \hline
 \multicolumn{7}{l}{$a$ - \citet{Pepin2006}.}\\   
 \multicolumn{7}{l}{$b$ - \citet{Meshik2020}.}\\
 \multicolumn{7}{l}{$c$ - \citet{Pepin1991}.}\\
 \multicolumn{7}{l}{$d$ - \citet{Ozima2002}.}\\
 \multicolumn{7}{l}{$e$ - \citet{Ott2012}.}\\
\end{tabular}
\label{table_nine}
\end{table}%

Given $^{244}$Pu's 80 Myr half-life and the expectation that early outgassing was efficient, we will assume that the fractionating escape took place after degassing of fissiogenic Xe (this is not an important restriction).
We can write this formally in terms of 9-isotope vectors:
\begin{equation}
\label{one}
 {\bf Air} = {\cal F} \left( {\bf U} + x{\bf Pu} + y{\bf I} \right) .
\end{equation}
The composition of U-Xe is recovered by de-fractionating the air and then subtracting off the radiogenic Xe
\begin{equation}
\label{two}
{\bf UXe} = {\cal F}^{-1}\!\left({\bf Air}\right) - x{\bf Pu} - y{\bf I} .
\end{equation}
For completeness we include $^{129}$I in Eqs \ref{one} and \ref{two},  
but as $^{129}$Xe maps uniquely to $y$, $^{129}$Xe can be treated separately. 

We have written fractionation as a general function ${\cal F}$.
Hydrodynamic escape can generate complicated fractionation patterns \citep{Hunten1987,Pepin1991,Dauphas2014},
but a simple expression is best here for illustration.
In the limit that the crossover mass $m_c$ (see Appendix A) exceeds the masses of all the escaping isotopes
($m_c\!>\!m_j,m_i$), fractionation in hydrodynamic escape reduces to a simple multiplier independent of $m_c$.
For specificity we express fractionations with respect to $^{130}$Xe.
\begin{equation}
\label{three}
 \mathrm{Air}_j = \exp{ \left( \eta \left(m_j-m_{130}\right)  \right) } \left( \mathrm{U}\mathrm{Xe}_j + x\mathrm{Pu}_j \right).
\end{equation}
The factor $\eta$ in Eq \ref{three} represents the magnitude of the fractionation.
Equation \ref{three} is trivially inverted to solve for a U-Xe:
\begin{equation}
\label{four}
 \mathrm{U}\mathrm{Xe}_j = \mathrm{Air}_j\exp{ \left( \eta \left(m_{130}-m_j\right)  \right) } - x\mathrm{Pu}_j
\end{equation}
This version of U-Xe is underdetermined, because the contribution $x$ of plutonium fission has not been specified.
We can set plausible bounds on the amount of Pu-Xe.  
The initial solar system abundance of $^{244}{\rm Pu}/^{238}{\rm U} = 0.0068$ \citep{Hudson1989},
when scaled to a bulk silicate Earth uranium abundance of 21 ppb 
implies that the initial bulk silicate Earth abundance of $^{244}$Pu was $3.0\times 10^{-10}$ [g/g]. 
The fractional yield for $^{244}\mathrm{Pu} \rightarrow\,\! ^{136}\mathrm{Xe}$ is $1.23\times 10^{-3}\times 0.056=6.8\times 10^{-5}$.
Hence the total production of plutoniogenic $^{136}$Xe on Earth has been $3.4\times 10^{11}$ moles.
This can be compared to the $1.4\times 10^{12}$ moles of $^{136}$Xe currently in the atmosphere.

\begin{figure}[htb] 
   \centering
   \includegraphics[width=5in]{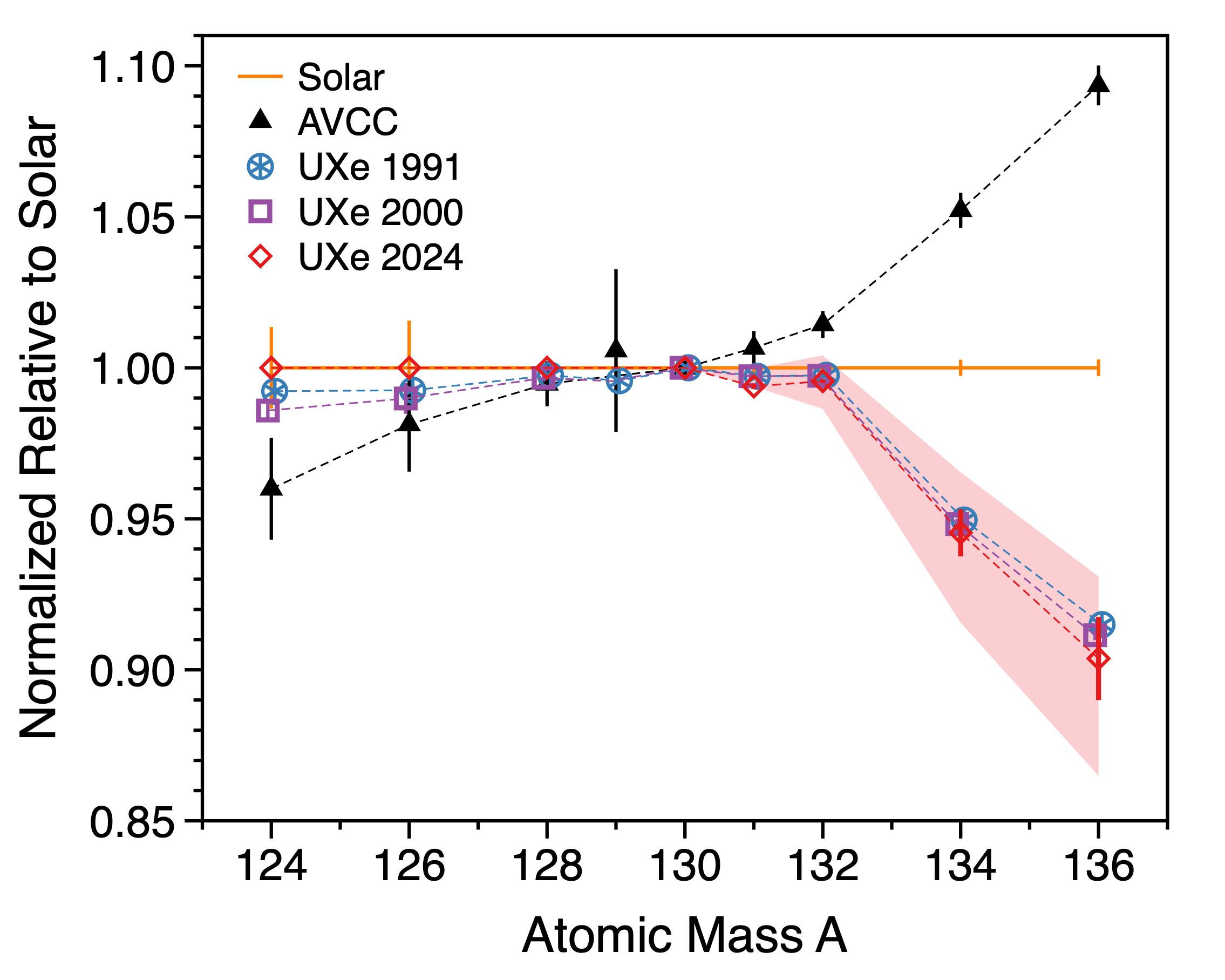} 
   \caption{Reconstructed isotopic compositions of U-Xe---Earth's birth Xe---compared to other selected xenons.
   The red shaded region is the range of possible U-Xe compositions consistent with plausible amounts of $^{244}$Pu fission Xe in air.
   This derivation does {\em not} take chondritic Xe into account.
   Standard U-Xe, quoted from \citet{Pepin1991} and \citet{Pepin2000}, 
   uses the composition of chondritic Xe to provide additional constraints.
   By construction U-Xe and SW-Xe are identical for light isotopes.
   The differences seen here stem from revisions of the solar wind. }
\label{fig:S1}
\end{figure}

Xenon's mass fractionation was accompanied by the loss of 65\%-90\% of the Xe that remained after impact erosion \citep{Zahnle2019}. 
This would have included Pu-Xe, as very little Pu-Xe remains in the mantle \citep{Kunz1998}.
With allowance for some early losses from
impact erosion---impact erosion probably decimated early Earth's noble gases \citep{Genda2005}, 
but most of those losses predated $^{244}\mathrm{Pu}$ decay---we expect Earth to have retained something like $\sim\!15^{+15}_{-7.5}\%$ of its plutoniogenic Xe.
This corresponds to $x \approx 8^{+8}_{-4}$, where $x$ is measured in units in which the abundance of $^{130}$Xe is 100.
The resulting range of plausible ``U-Xe'' compositions is indicated in Figure \ref{fig:S1} by pink shading.
Note that this simple argument makes no reference to chondritic Xe.

\citet{Pepin1991} originally proposed that U-Xe was the true solar Xe, which implied that putative solar Xe \citep[e.g., ``SUCOR,''][]{Ozima2002}
was U-Xe polluted with exotic Xe.
Schematically, 
  \begin{equation}
 \label{five_a}
{\bf SW} = \mathbf{UXe} + z{\bf HXe} ,
\end{equation}
in which $z$ is an optimizable parameter, and
the pollutant---H-Xe---closely resembles Xe-H with $^{131}$Xe mostly removed.
This U-Xe is plotted on Figure \ref{fig:S1} as ``U-Xe 1991.'' 
Later, with SW-Xe firmly established, \citet{Pepin2000} revised the derivation of U-Xe,
which in heuristic form can be written
 \begin{equation}
 \label{five}
\mathbf{UXe} = {\bf SW} - z{\bf HXe}.
\end{equation}
In the new derivation, H-Xe was redefined to include only $^{134}$Xe and $^{136}$Xe.
The revised U-Xe is plotted on Figure \ref{fig:S1} as ``U-Xe 2000.''

\begin{figure}[htb]
 \centering
 \begin{minipage}[c]{0.6\textwidth}
   \centering
  \includegraphics[width=1\textwidth]{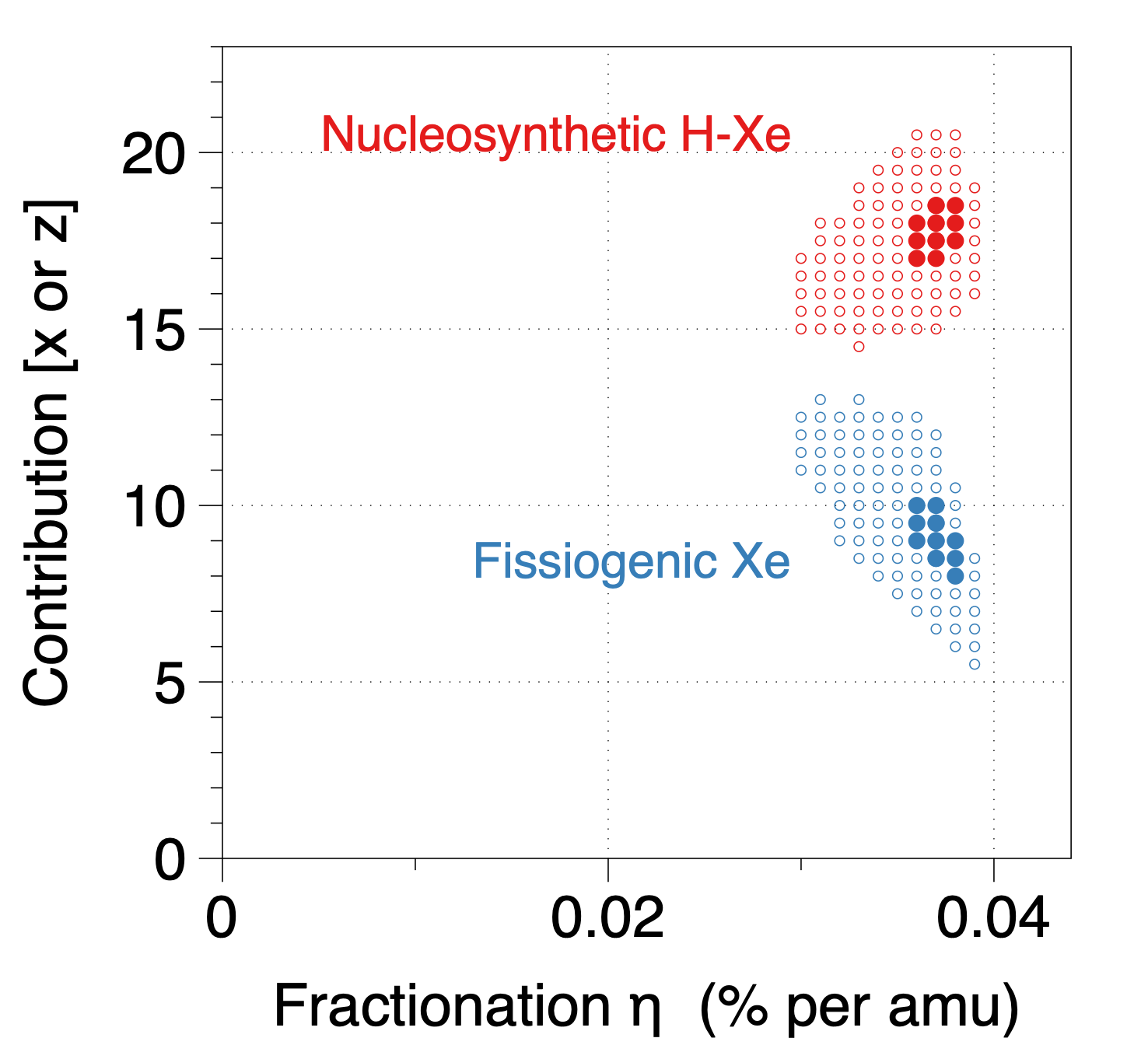} 
 \end{minipage}
 \begin{minipage}[c]{0.38\textwidth}
   \centering
  \caption{Reconstructing U$^{\prime}$-Xe.
    Beginning with solar xenon, a quantity $x$ of $^{244}$Pu fissiogenic Xe is added 
     and a quantity $z$ of Xe-H is subtracted
   ($x$ and $z$ are in units where $^{130}\mathrm{Xe} \!=\!100$).
    The sum is then mass fractionated by $\eta$ to generate air's Xe ($\eta$ is in units of $\%$ per amu). 
    Values of $x$, $z$, and $\eta$ that give acceptable fits are plotted.
     The umbrae of filled disks have $\chi^2\!<\!1$.
    The penumbrae of open circles have $\chi^2\!<\!3.5$. 
    The best fits are used to create a U$^{\prime}$-Xe using Eq \ref{five},
    which is plotted on Figure \ref{fig:S1}.
    }
\label{fig:S2}
 \end{minipage}
\end{figure}

In their details, Pepin's derivations are rather obscure.
We therefore independently performed a simple Monte Carlo simulation by taking random values from Gaussian distributions defined by
the quoted uncertainties (one $\sigma$ errors are listed in Table \ref{table_nine}).  
For H-Xe we followed \citep{Ott2012}.
Misfits were defined as
 \begin{equation}
 \label{six}
\Delta_j = \mathrm{Air}_j \exp{ \left( \eta \left(m_{130}-m_j\right)  \right) } - \mathrm{SW}_j - x\mathrm{Pu}_j  + z\mathrm{H}\mathrm{Xe}_j .
\end{equation}
From these we computed $\chi^2$ values for each realization  
 \begin{equation}
 \label{seven}
\chi^2 = \sum_{j\neq 129} {  \Delta_j^2 \over \sigma^2_{Air_j} + \sigma^2_{SW_j} + x^2\sigma^2_{Pu_j} + z^2\sigma^2_{H\!Xe_j}}.
\end{equation}
Results are shown in Figure \ref{fig:S2}.
For this particular numerical experiment,
we find that $x=9\pm 3$, $z=17.5\pm 2.5$, and $\eta=0.037\pm0.002$. 
This U-Xe is plotted on Figure \ref{fig:S1} as ``U-Xe 2024.''
It is very similar to the official U-Xe, differing mostly in providing more room for uncertainties.

\end{document}